\documentclass[a4paper]{article}

\usepackage[margin=1in]{geometry}

\usepackage[T1]{fontenc}
\newcommand{\changefont}[3]{
\fontfamily{#1} \fontseries{#2} \fontshape{#3} \selectfont}

\changefont{ptm}{m}{n}

\usepackage{setspace} 
\usepackage{graphicx}    

\usepackage{amsfonts}
\usepackage{amssymb}
\usepackage{graphics}
\usepackage{mathrsfs}
\usepackage{color}

\newtheorem{theorem}{Theorem}[section]

\newtheorem{lemma}{Lemma}[section]

\long\def\symbolfootnote[#1]#2{\begingroup%
\def\thefootnote{\fnsymbol{footnote}}\footnote[#1]{#2}\endgroup} 

\begin{document}

\begin{center}
\Large \textbf{Homoclinical structure of SICNNs with rectangular input currents}
\end{center}

\begin{center}
\normalsize \textbf{Mehmet Onur Fen$^{1,}\symbolfootnote[1]{Corresponding Author. E-mail: monur.fen@gmail.com, Tel: +90 312 585 02 17}$, Fatma Tokmak Fen$^2$} \\
\vspace{0.2cm}
\textit{\textbf{$^1$Basic Sciences Unit, TED University, 06420 Ankara, Turkey}}

\vspace{0.1cm}
\textit{\textbf{$^2$Department of Mathematics, Gazi University, 06500 Ankara, Turkey}}
\vspace{0.1cm}
\end{center}

\vspace{0.3cm}

\begin{center}
\textbf{Abstract}
\end{center}

\vspace{-0.2cm}

\noindent\ignorespaces

Shunting inhibitory cellular neural networks (SICNNs) with continuous as well as discontinuous external inputs are investigated. The descriptions of homoclinic and heteroclinic motions are provided in the functional sense for the multidimensional dynamics of SICNNs, and it is demonstrated that the networks under investigation exhibit such motions. Homoclinic and heteroclinic outputs that are asymptotic to quasi-periodic outputs are illustrated.

\vspace{0.2cm}
 
\noindent\ignorespaces \textbf{Keywords:} Shunting inhibitory cellular neural networks; Homoclinic and heteroclinic outputs; Rectangular input currents; Stable and unstable sets; Quasi-periodic outputs

\vspace{0.6cm}


\section{Introduction}

Starting with the studies Poincar\'{e} \cite{Andersson94} the investigation of homoclinic and heteroclinic motions has been one of the attractive topics among scientists who study nonlinear dynamics.
The presence of homoclinic motions is an effective tool for the demonstration of chaotic behavior since any neighborhood of a structurally stable Poincar\'{e} homoclinic orbit admits nontrivial hyperbolic sets that contain a countable number of saddle periodic orbits and continuum of non-periodic Poisson stable orbits \cite{Gonchenko96}-\cite{Smale65}. Besides, heteroclinic motions are also crucial for chaotic dynamics \cite{Bertozzi88,Chacon95}.

The present study is devoted to the investigation of homoclinic and heteroclinic motions generated by shunting inhibitory cellular neural networks (SICNNs), which is a class of biologically inspired cellular neural networks introduced by Bouzerdoum and Pinter \cite{Bouzerdoum93}. This type of neural networks have been extensively applied in psychophysics, speech, perception, robotics, adaptive pattern recognition, vision and image processing \cite{bouzer4}-\cite{Pinter89}.

The model of SICNNs in its most original form is as follows \cite{Bouzerdoum93}. Consider a two dimensional grid of processing cells arranged into $m$ rows and $n$ columns, and let $C_{ij}$ denote the cell at the $(i,j)$ position of the lattice, where $i=1,2,\ldots,m$ and $j=1,2,\ldots, n.$ In SICNNs, neighboring cells exert mutual inhibitory interactions of the shunting type. Define  the $r-$neighborhood of the cell $C_{ij}$   as
\begin{eqnarray*} 
\begin{array}{l} 
N_{r}(i,j)=\{C_{hl}: \max(|h-i|,|l-j|)\leq r,  \ 1\leq h\leq m,  \ 1\leq l \leq n \}.
\end{array}
\end{eqnarray*} 
The dynamics of a cell $C_{ij}$ are described by the following nonlinear ordinary differential equation:
\begin{eqnarray} \label{eq:1}
\displaystyle \frac{dx_{ij}}{dt}=-a_{ij}x_{ij}-\sum_{C_{hl}\in N_{r}(i,j)} C_{ij}^{hl}f(x_{hl}(t))x_{ij}  + \mathscr{L}_{ij}(t), 
\end{eqnarray}
where $x_{ij}$ is the activity of the cell $C_{ij};$ $\mathscr{L}_{ij}(t)$ is the external input to the cell $C_{ij};$ the constant $a_{ij}>0$ represents the passive decay rate of the cell activity; $C_{ij}^{hl}\geq 0$ is the coupling strength of post-synaptic activity of the cell $C_{hl}$ transmitted to the cell $C_{ij};$ the positive and continuous function $f(x_{hl})$ represents the output or firing rate of the cell $C_{hl}.$

The existence and stability of periodic, almost-periodic and anti-periodic solutions of SICNNs have been widely studied in the literature \cite{Peng09}-\cite{Liu15}. Moreover, chaos in SICNNs were considered in the papers \cite{Ge06}-\cite{Akh12}. However, to the best of our knowledge, the existence of homoclinic and heteroclinic solutions of SICNNs has not been investigated before in the literature. Motivated by the lack of mathematical methods for the investigation of homoclinic and heteroclinic motions in SICNNs, we propose the results of the present study. Both continuous and discontinuous external inputs in rectangular form are utilized in our model. 

The generation of homoclinic and heteroclinic motions in systems of differential equations by means of discontinuous perturbations was first realized by Akhmet \cite{Akh2} on the basis of functional spaces. The paper \cite{Akh2} was concerned with the presence of homoclinic motions embedded in the chaotic attractor of a relay system (see \cite{Akh6,Book16} for details about relay systems). Similar results for impulsive differential equations were obtained in the studies \cite{Akh7,Fen16}, and the formation of homoclinic and heteroclinic motions in economic models was considered within the scope of the paper \cite{Akh10}. Based on the definitions given in \cite{Akh2}, in the present study, we provide the descriptions of homoclinic and heteroclinic motions for the multidimensional dynamics of SICNNs in the functional sense, and rigorously prove that the discontinuous inputs give rise to the generation of homoclinic and heteroclinic motions.

One of the techniques for the confirmation of chaos in neural networks is to determine homoclinic and heteroclinic motions. A geometric construction of a transversal homoclinic orbit for a nonlinear neuron model with time delays was provided by Chen \cite{Chen09} in order to show the existence of chaos. Moreover, Li et al. \cite{Li14} proved the existence of a saddle periodic orbit in the dynamics of a small Hopfield neural network with a memristive synaptic weight, and verified the existence of a hyperchaotic invariant set via a homoclinic intersection of its stable and unstable manifolds by using the Smale-Birkhoff homoclinic theorem. Zou et al. \cite{Zou93}, on the other hand, presented a method for finding homoclinic and heteroclinic motions in the three-cell autonomous cellular neural network, which has chaotic behavior similar to that of the double scroll attractor in a certain parameter range.  In the study \cite{Zou93}, homoclinic motions that are asymptotic to equilibria are considered. However, in this paper, we take into account a larger class of homoclinic and heteroclinic motions, and illustrate the presence of such motions which are asymptotic to quasi-periodic ones in Section \ref{example_sec}.

The rest of the paper is organized as follows. In the next section the model of SICNNs with rectangular input currents (RICs) is introduced, and some results concerning the bounded solutions are mentioned. 
Section \ref{homSICNN_sec3} is devoted to the descriptions of homoclinic and heteroclinic motions of SICNNs as well as the main result of the present study. An illustrative example, which support the theoretical results, is provided in Section \ref{example_sec}. Finally, some concluding remarks are given in Section \ref{SICNNhom_conc}.

\section{The model}

We will take into account the external input $\mathscr{L}_{ij}(t)$ in SICNN (\ref{eq:1}) as the sum of the continuous external input $L_{ij}(t)$ and the RIC $P_{ij}(t,\zeta).$ More precisely, the main object of investigation is the following SICNN, 
\begin{eqnarray} \label{mainmodel}
\displaystyle \frac{dx_{ij}}{dt}=-a_{ij}x_{ij}-\sum_{C_{hl}\in N_{r}(i,j)} C_{ij}^{hl}f(x_{hl}(t))x_{ij}  + L_{ij}(t) + P_{ij}(t,\zeta). 
\end{eqnarray}
In model (\ref{mainmodel}), for each $i=1,2,\ldots,m$ and $j=1,2,\ldots,n,$ the input currents $P_{ij}(t,\zeta)$ are defined through the equations $P_{ij}(t,\zeta)=\zeta^{ij}_k$ for $t\in (\theta_{k}, \theta_{k+1}],$ $k\in\mathbb Z,$ such that the sequence $\left\{\theta_k\right\}_{k \in \mathbb Z}$ of discontinuity moments is strictly increasing, $\left|\theta_k\right| \to \infty$ as $ \left|k\right|\to\infty,$ and $\zeta=\left\{\zeta_k\right\}_{k \in \mathbb Z}$ is a sequence generated by the map 
\begin{eqnarray}\label{discrete_map}
\zeta_{k+1}= F(\zeta_k), 
\end{eqnarray}
where $F:\Lambda \to \Lambda$ is a continuous function and $\Lambda$ is a bounded subset of $\mathbb R^{mn}.$  
Our purpose is to prove the existence of homoclinic and heteroclinic motions in the dynamics of SICNN (\ref{mainmodel}) in the case that the map (\ref{discrete_map}) admits homoclinic and heteroclinic orbits.

The dynamics of SICNNs with RICs were also investigated in the paper \cite{Akh8}. It was found in \cite{Akh8} that if the discontinuity moments of the RICs change chaotically, then the network possesses chaos with infinitely many almost periodic motions in basis. Homoclinic and heteroclinic motions were not investigated in \cite{Akh8}, and the novelty of the present study is the investigation of such motions in the dynamics of SICNNs. Another difference is that the discontinuity moments of the RICs are generated by a discrete chaotic map in \cite{Akh8}, while the values of the RICs in the intervals of constancy are generated by a discrete map, which possesses homoclinic and heteroclinic orbits. Besides, it was mentioned in \cite{Akh8} that the discontinuity moments of the RICs can be comprehended as spike moments, and it is reasonable to use networks with this type of inputs in modeling biological neural systems.

In an electrical equivalent circuit of a cell of SICNNs, the synaptic conductance of each inhibitory channel is proportional to the firing rate of the cell controlling it, and the shunting conductance of the cell is the sum of the synaptic conductances of individual inhibitory channels \cite{Bouzerdoum93}. Since both continuous and discontinuous external inputs are utilized, one may design an electrical equivalent circuit of the network (\ref{mainmodel}) by properly coupling appropriate pulse generating circuits \cite{Tyagi76,Ping03} with the one mentioned in \cite{Bouzerdoum93}.

Throughout the paper, $\mathbb R$ and $\mathbb Z$ will stand for the sets of real numbers and integers, respectively, and the norm $\left\|u\right\|=\displaystyle \max_{(i,j)} \left|u_{ij}\right|$ will be used, where $u=\left\{u_{ij}\right\} = (u_{11},\ldots,u_{1n}, \ldots, u_{m1} \ldots,u_{mn}) \in \mathbb R^{m\times n}.$  

The following conditions are required.
\begin{itemize}
\item[\textbf{(C1)}] There exists a positive number $L_f$ such that  $\left|f(s_1)-f(s_2)\right| \le L_f \left|s_1-s_2\right|$ for all $s_1,$ $s_2 \in \mathbb R;$
\item[\textbf{(C2)}] There exist positive numbers $M_f$ and $M_{ij}$ such that $\displaystyle \sup_{s \in \mathbb R} \left|f(s)\right| \le M_f$ and $\displaystyle \sup_{t \in \mathbb R} \left|L_{ij}(t)\right| \le M_{ij}$ for each $i=1,2,\ldots,m,$ $j=1,2,\ldots,n.$
\end{itemize}

Suppose that the inequality $M_f \displaystyle \max_{(i,j)}\frac{  \sum_{C_{hl} \in N_r(i,j)} C_{ij}^{hl}}{a_{ij}} < 1$ is valid. Let us denote $\displaystyle \gamma = \min_{(i,j)} a_{ij},$ $\delta=\displaystyle \max_{(i,j)} \sum_{C_{hl} \in N_r(i,j)} C_{ij}^{hl}$ and $K_0= \displaystyle \frac{\displaystyle \max_{(i,j)} \frac{M_{ij}+M_F}{a_{ij}}}{ 1- M_f\displaystyle \max_{(i,j)}\frac{ \sum_{C_{hl} \in N_r(i,j)} C_{ij}^{hl}}{a_{ij}} },$ 
where $M_F=\displaystyle \sup_{\xi \in \Lambda} \left\|F(\xi)\right\|.$ It is worth noting that the number $\gamma$ is positive and $\delta$ is nonnegative since the numbers $a_{ij}$ are positive and the coupling strengths $C_{ij}^{hl}$ are nonnegative.

The following condition is also needed.
\begin{itemize}
\item[\textbf{(C3)}] $ \gamma - (M_f+L_fK_0) \delta  > 0.$
\end{itemize}

Since the model (\ref{mainmodel}) satisfies the Lipschitz condition in the intervals of constancy of $P_{ij}(t,\zeta),$ the local existence and uniqueness of solutions are valid, and at the discontinuity moments, one can use the continuity condition of solutions to proceed for the next interval of constancy of $P_{ij}(t,\zeta).$ Therefore, the local existence and uniqueness of solutions as well as their continuation to the maximal interval of existence are assured. The reader is referred to the books \cite{Filippov88,Akhmet2011book} for more information about differential equations with discontinuous right hand side.

If the conditions $(C1)-(C3)$ are valid, then the technique mentioned in the paper \cite{Akh9} can be used to prove that for each solution $\zeta=\left\{\zeta_k\right\}_{k\in\mathbb Z}$ of (\ref{discrete_map}), there exists a unique bounded on $\mathbb R$ solution $\phi_{\zeta}(t) = \left\{\phi_{\zeta}^{ij}(t)\right\},$ $i=1,2,\ldots,m,$ $j=1,2,\ldots,n,$ of (\ref{mainmodel}), which satisfies the following relation
\begin{eqnarray} \label{int_relation}
\phi^{ij}_{\zeta}(t)=\displaystyle - \int_{-\infty}^t e^{-a_{ij} (t-s)}  \bigg[ \sum_{C_{hl}\in N_r(i,j)}  C_{ij}^{hl}  f(\phi^{hl}_{\zeta}(s))\phi^{ij}_{\zeta}(s) - L_{ij}(s) - P_{ij}(s,\zeta) \bigg] ds 
\end{eqnarray}
such that $\displaystyle \sup_{t\in \mathbb R} \left\|\phi_{\zeta}(t)\right\| \le K_0.$ Moreover, under the same conditions, it can be verified in a very similar way to Theorem 2 \cite{Huang03} that for a fixed solution $\zeta=\left\{\zeta_k\right\}_{k \in\mathbb Z}$ of (\ref{discrete_map}) all other solutions of SICNN (\ref{mainmodel}) converge exponentially to the unique bounded on $\mathbb R$ solution $\phi_{\zeta}(t).$

\section{Homoclinic and heteroclinic motions} \label{homSICNN_sec3}

Let us describe the stable, unstable and hyperbolic sets as well as the homoclinic and heteroclinic motions for both SICNN (\ref{mainmodel}) and the discrete map (\ref{discrete_map}). These definitions are adapted from the papers \cite{Akh2,Akh7}. 

Suppose that $\Theta$ denotes the set of all sequences $\zeta=\left\{\zeta_k\right\}_{k\in\mathbb Z} \subset \Lambda$ obtained by (\ref{discrete_map}). The stable set of a sequence $\zeta\in\Theta$ is defined as 
\begin{eqnarray*} \label{stable_set}
W^s(\zeta)= \left\{ \eta \in \Theta \ | \ \left\|\eta_k-\zeta_k\right\|\to 0 ~\textrm{as}~ k\to  \infty  \right\},
\end{eqnarray*}
and the unstable set of $\zeta$ is 
\begin{eqnarray*} \label{unstable_set}
W^u(\zeta)= \left\{ \eta \in \Theta \ | \ \left\|\eta_k-\zeta_k\right\|\to 0 ~\textrm{as}~ k\to  -\infty  \right\}.
\end{eqnarray*}
The set $\Theta$ is called hyperbolic if for each $\zeta \in \Theta$ the stable and unstable sets of $\zeta$ contain at least one element different from $\zeta.$ A sequence $\eta \in \Theta$ is homoclinic to another sequence $\zeta \in \Theta$ if $\eta \in W^s(\zeta) \cap W^u(\zeta).$ Moreover, $\eta \in \Theta$ is heteroclinic to the sequences $\zeta^1 \in \Theta,$ $\zeta^2 \in \Theta,$ $\eta \neq \zeta^1,$ $\eta \neq \zeta^2,$ if  $\eta \in W^s(\zeta^1) \cap W^u(\zeta^2).$

On the other hand, let $\mathscr{A}$ be the set consisting of all bounded on $\mathbb R$ solutions $\phi_{\zeta}(t),$ $\zeta \in \Theta,$ of SICNN (\ref{mainmodel}). 
A bounded solution $\phi_{\eta}(t) \in \mathscr{A}$ belongs to the stable set $W^s(\phi_{\zeta}(t))$ of  $\phi_{\zeta}(t) \in \mathscr{A}$ if $\left\|\phi_{\eta}(t)-\phi_{\zeta}(t)\right\|\to 0$ as $t\to  \infty.$ Besides,  $\phi_{\eta}(t)$ is an element of the unstable set $W^u(\phi_{\zeta}(t))$ of  $\phi_{\zeta}(t)$ provided that  $\left\|\phi_{\eta}(t)-\phi_{\zeta}(t)\right\|\to 0$ as $t\to -\infty.$ 

We say that $\mathscr{A}$ is  hyperbolic if for each  $\phi_{\zeta}(t) \in \mathscr{A}$ the sets $W^s(\phi_{\zeta}(t))$ and $W^u(\phi_{\zeta}(t))$ contain at least one element different from $\phi_{\zeta}(t).$ A solution $\phi_{\eta}(t)\in \mathscr{A}$ is homoclinic to another solution $\phi_{\zeta}(t) \in \mathscr{A}$ if $\phi_{\eta}(t) \in W^s(\phi_{\zeta}(t)) \cap W^u(\phi_{\zeta}(t)),$ and 
$\phi_{\eta}(t)\in \mathscr{A}$ is heteroclinic to the bounded solutions $\phi_{\zeta^1}(t),$ $\phi_{\zeta^2}(t)\in \mathscr{A},$ $\phi_{\eta}(t) \neq \phi_{\zeta^1}(t),$ $\phi_{\eta}(t) \neq \phi_{\zeta^2}(t),$ if $\phi_{\eta}(t) \in W^s(\phi_{\zeta^1}(t)) \cap W^u(\phi_{\zeta^2}(t)).$

The relations between the stable and unstable sets of the network (\ref{mainmodel}) and the discrete map (\ref{discrete_map}) are respectively provided in Lemma \ref{lemma1} and Lemma \ref{lemma2}.

\begin{lemma}\label{lemma1} 
Assume that the conditions $(C1)-(C3)$ are valid, and let $\zeta$ and $\eta$ be elements of $\Theta.$ If $\eta\in W^{s}(\zeta),$ then $\phi_{\eta}(t)\in W^{s}(\phi_{\zeta}(t)).$  
\end{lemma}

\noindent \textbf{Proof.} 
Fix an arbitrary positive number $\epsilon,$ and let $\alpha$ be a real number such that $$\alpha\ge 1+\displaystyle\frac{1}{\gamma-\left(M_f+L_f K_0\right)\delta}.$$  Since $\eta=\left\{\eta_k\right\}_{k\in\mathbb Z}$ is an element of $W^{s}(\zeta),$ there exists an integer $k_0$ such that $\left\|\eta_k-\zeta_k\right\|<\displaystyle\frac{\epsilon}{\alpha}$ for all $k\ge k_0.$ 

According to (\ref{int_relation}), the bounded solutions $\phi_{\eta}(t)$ and $\phi_{\zeta}(t)$ satisfy the following relation
\begin{eqnarray*} 
&& \phi_{\eta}^{ij}(t) - \phi_{\zeta}^{ij}(t) = -\displaystyle \int_{-\infty}^{t} e^{-a_{ij}(t-s)} \Big[\displaystyle \sum_{C_{hl}\in N_{r}(i,j)}C_{ij}^{hl} f \big( \phi_{\eta}^{hl} (s) \big) \phi_{\eta}^{ij}(s) -P_{ij}(s,\eta)\\
&&- \displaystyle \sum_{C_{hl}\in N_{r}(i,j)} C_{ij}^{hl} f \big( \phi_{\zeta}^{hl} (s) \big) \phi_{\zeta}^{ij}(s) +P_{ij}(s,\zeta)  \Big] ds. \\
\end{eqnarray*}
Using the inequality  $\left| P_{ij}(t,\eta)-P_{ij}(t,\zeta)\right| <\displaystyle \frac{\epsilon}{\alpha},$   $t \in (\theta_{k_0},\infty),$ one can obtain on the same interval that
\begin{eqnarray*}
\begin{array}{l}
\displaystyle  \big|\phi_{\eta}^{ij}(t) - \phi_{\zeta}^{ij}(t)\big| \le 2(M_f K_0 \displaystyle \sum_{C_{hl}\in N_{r}(i,j)} C_{ij}^{hl} +M_F)\displaystyle \int_{-\infty}^{\theta_{k_0}}  e^{-a_{ij} (t-s)} ds  \\ 
+ \displaystyle \int_{\theta_{k_0}}^{t}  e^{-a_{ij} (t-s)} M_f\displaystyle \sum_{C_{hl}\in N_{r}(i,j)} C_{ij}^{hl} \displaystyle  \big|\phi_{\eta}^{ij}(s) - \phi_{\zeta}^{ij}(s)\big| ds\\
+ \displaystyle \int_{\theta_{k_0}}^{t}  e^{-a_{ij} (t-s)} L_f K_0\displaystyle \sum_{C_{hl}\in N_{r}(i,j)} C_{ij}^{hl} \displaystyle  \big|\phi_{\eta}^{hl}(s) - \phi_{\zeta}^{hl}(s)\big| ds+\displaystyle \int_{\theta_{k_0}}^{t}  e^{-a_{ij}(t-s)}\frac{\epsilon}{\alpha} ds.
\end{array}
\end{eqnarray*}
Therefore, we have for $t>\theta_{k_0}$ that
\begin{eqnarray} \label{hom_proof_ineq1}
\begin{array}{l}
\displaystyle \big\|\phi_{\eta}(t) - \phi_{\zeta}(t)\big\| \le 2 \max_{(i,j)}\frac{1}{a_{ij}}\left(M_f K_0 \displaystyle \sum_{C_{hl}\in N_{r}(i,j)} C_{ij}^{hl} +M_F\right) e^{-\gamma (t-\theta_{k_0})} +\frac{\epsilon}{\gamma\alpha}\left(1-e^{-\gamma (t-\theta_{k_0})}\right)  \\
+(M_f+L_f K_0)\displaystyle\max_{(i,j)}\displaystyle \sum_{C_{hl}\in N_{r}(i,j)} C_{ij}^{hl}  \int_{\theta_{k_0}}^{t} e^{-\gamma (t-s)} \big\|\phi_{\eta}(s) - \phi_{\zeta}(s)\big\|ds.
\end{array}
\end{eqnarray}
Define the function $v(t)=e^{\gamma t} \left\|  \phi_{\eta}(t) - \phi_{\zeta}(t)  \right\|$ and let $$R_0=2\displaystyle  \max_{(i,j)}\frac{1}{a_{ij}}\left(M_f K_0 \displaystyle \sum_{C_{hl}\in N_{r}(i,j)} C_{ij}^{hl}+M_F \right)-\frac{\epsilon}{\gamma\alpha}.$$
The inequality (\ref{hom_proof_ineq1}) yields
\begin{eqnarray*} 
v(t)\le R_0 e^{\gamma \theta_{k_0}}+\displaystyle\frac{\epsilon}{\gamma\alpha}e^{\gamma t}+(M_f+L_f K_0)\delta \int_{\theta_{k_0}}^{t} v(s) ds, \ t>\theta_{k_0}.
\end{eqnarray*}
Applying the Gronwall's Lemma \cite{Corduneanu71} to the last inequality one can verify that
\begin{eqnarray*}
&v(t) &\le \displaystyle\frac{\epsilon}{\gamma\alpha}e^{\gamma t}+R_0e^{(M_f+L_f K_0) \delta (t-\theta_{k_0})} e^{\gamma \theta_{k_0}}\\
&&+\displaystyle\frac{\epsilon (M_f+L_f K_0) \delta}{\gamma\alpha \left[\gamma-(M_f+L_f K_0) \delta\right]}e^{\gamma t}\left(1-e^{-\left[\gamma-(M_f+L_f K_0) \delta\right](t-\theta_{k_0})}\right).
\end{eqnarray*}
Multiplying both sides of the last inequality by $e^{-\gamma t}$ we obtain that  
\begin{eqnarray*}
\left\|\phi_{\eta}(t) - \phi_{\zeta}(t) \right\| <\displaystyle\frac{\epsilon}{\alpha \left[\gamma-(M_f+L_f K_0)\delta\right]}+R_0 e^{-\left[\gamma-(M_f+L_f K_0)\delta\right](t-\theta_{k_0})}, \ t>\theta_{k_0}.
\end{eqnarray*}
Now, let $E$ be a positive number such that
$
E\ge\displaystyle\frac{1}{\gamma-(M_f+L_f K_0)\delta} \ln\left(\frac{R_0 \alpha}{\epsilon}\right).
$
Hence, for $t\ge E+\theta_{k_0},$ we have
\begin{eqnarray*}
\left\|\phi_{\eta}(t)-\phi_{\zeta}(t)\right\|<\displaystyle\frac{\epsilon}{\alpha}\Big(1+\frac{1}{\gamma-(M_f+L_f K_0)\delta}\Big)\le \epsilon. 
\end{eqnarray*}
Consequently, $\phi_{\eta}(t)$ belongs to $W^{s}(\phi_{\zeta}(t)).$ $\square$

\begin{lemma}\label{lemma2} 
Assume that the conditions $(C1)-(C3)$ are valid, and let $\zeta$ and $\eta$ be elements of $\Theta.$ If $\eta\in W^{u}(\zeta),$ then $\phi_{\eta}(t)\in W^{u}(\phi_{\zeta}(t)).$  
\end{lemma}

\noindent \textbf{Proof.} 
Fix an arbitrary positive number $\epsilon$, and let $\beta$ be a real number satisfying $$\beta>\displaystyle \frac{1}{\gamma} \bigg(1-(M_f+ L_f K_0)\max_{(i,j)}\frac{\sum_{C_{hl}\in N_{r}(i,j)} C_{ij}^{hl}}{a_{ij}}  \bigg)^{-1} .$$ 

Because the sequence $\eta=\left\{\eta_k\right\}_{k\in\mathbb Z}$ belongs to the unstable set $W^{u}(\zeta)$ of $\zeta=\left\{\zeta_k\right\}_{k\in\mathbb Z},$ there exists an integer $k_0$ such that $\left\|\eta_k-\zeta_k\right\|<\displaystyle\frac{\epsilon}{\beta}$ for all $k\le k_0.$ In this case, for each $i$ and $j,$ one can confirm that $\left| P_{ij}(t,\eta)-P_{ij}(t,\zeta) \right|<\displaystyle \frac{\epsilon}{\beta},$ $t\le\theta_{k_0+1}.$

Making use of the relation
\begin{eqnarray*} 
&& \phi_{\eta}^{ij}(t) - \phi_{\zeta}^{ij}(t) = -\displaystyle \int_{-\infty}^{t} e^{-a_{ij}(t-s)} \Big[\displaystyle \sum_{C_{hl}\in N_{r}(i,j)}C_{ij}^{hl} f \big( \phi_{\eta}^{hl} (s) \big) \phi_{\eta}^{ij}(s) -P_{ij}(s,\eta)\\
&&- \displaystyle \sum_{C_{hl}\in N_{r}(i,j)} C_{ij}^{hl} f \big( \phi_{\zeta}^{hl} (s) \big) \phi_{\zeta}^{ij}(s) +P_{ij}(s,\zeta)  \Big] ds, 
\end{eqnarray*}
it can be verified for $t\le\theta_{k_0+1}$ that
\begin{eqnarray*}
\displaystyle  \big|\phi_{\eta}^{ij}(t) - \phi_{\zeta}^{ij}(t)\big| < \frac{\epsilon}{\beta a_{ij}}+(M_f+ L_f K_0)\frac{\sum_{C_{hl}\in N_{r}(i,j)} C_{ij}^{hl} }{a_{ij}}  \sup_{t\le \theta_{k_0+1}}\left\|\phi_{\eta}(t) - \phi_{\zeta}(t)\right\|. 
 \end{eqnarray*}
The last inequality implies that
\begin{eqnarray*}
 \sup_{t\le \theta_{k_0+1}}\left\|\phi_{\eta}(t) - \phi_{\zeta}(t)\right\|  \le \frac{\epsilon}{\gamma \beta} \bigg(1-(M_f+ L_f K_0)\max_{(i,j)}\frac{\sum_{C_{hl}\in N_{r}(i,j)} C_{ij}^{hl}}{a_{ij}}  \bigg)^{-1} < \epsilon.
 \end{eqnarray*}
Hence, $\displaystyle \lim_{t \to -\infty} \left\|\phi_{\eta}(t) - \phi_{\zeta}(t)\right\| = 0.$ $\square$

The following theorem can be proved using the results of Lemma \ref{lemma1} and Lemma \ref{lemma2}.

\begin{theorem}\label{main_theorem}
Under the conditions $(C1)-(C3),$ the following assertions are valid.
\begin{enumerate}
\item[(i)] If $\eta \in \Theta$ is homoclinic to $\zeta \in \Theta,$ then $\phi_{\eta}(t) \in \mathscr{A}$ is homoclinic to $\phi_{\zeta}(t) \in \mathscr{A};$
\item[(ii)] If $\eta \in \Theta$ is heteroclinic to $\zeta^1,$ $\zeta^2 \in \Theta,$ then $\phi_{\eta}(t) \in \mathscr{A}$ is heteroclinic  to $\phi_{\zeta^1}(t),$ $\phi_{\zeta^2}(t) \in \mathscr{A};$
\item[(iii)] If $\Theta$ is hyperbolic, then the same is true for $\mathscr{A}.$
\end{enumerate}
\end{theorem}

The next section is devoted to an illustrative example.

\section{An Example} \label{example_sec}

Let us consider the SICNN
\begin{eqnarray} \label{ex1}
\displaystyle \frac{dx_{ij}}{dt}=-a_{ij}x_{ij}-\sum_{C_{hl}\in N_{1}(i,j)} C_{ij}^{hl}f(x_{hl}(t))x_{ij}  + L_{ij}(t) + P_{ij}(t,\zeta), 
\end{eqnarray}
where $i,j=1,2,3,$ $f(s)=\displaystyle \frac{\arctan(s^2)}{9},$ $\theta_k=\displaystyle \frac{7k}{2},$ $k\in\mathbb Z,$ $L_{11}(t)=0.4\sin(2\pi t)+0.6\cos(\sqrt{5} t),$ $L_{12}(t)=0.4 \cos (\sqrt{2} t)-0.2\cos t,$ $L_{13}(t)=0.5\sin(3t) +0.3\sin(2\pi t),$ $L_{21}(t)=0.1\sin(3\pi t)-0.6\cos(\sqrt{2}t),$ $L_{22}(t) =0.2 \sin (\pi t) + 0.1 \cos(2t),$ $L_{23}(t)=0.1\cos(2\sqrt{3}t)+0.7\cos(5\pi t),$ $L_{31}(t)=0.4\sin(\pi t/2)+0.5\cos t,$ $L_{32}(t)=0.7 \sin (3 t) - 0.1 \sin(\sqrt{2}t),$  $L_{33}(t)=0.5 \sin (\sqrt{3} t) + 0.2 \sin(5t),$  
\[
\left( 
{\begin{array}{ccc}
a_{11} & a_{12} & a_{13} \\       
a_{21} & a_{22} & a_{23} \\ 
a_{31} & a_{32} & a_{33}     
\end{array}} 
\right)
=
\left( 
{\begin{array}{ccc}
0.4 & 0.6 & 0.5 \\       
0.7 & 0.3 & 0.8 \\ 
0.4 & 0.9 & 0.6     
\end{array}} 
\right),
\] and
\begin{eqnarray*}
\left( 
{\begin{array}{ccc}
C_{ij}^{11} & C_{ij}^{12} & C_{ij}^{13} \\       
C_{ij}^{21} & C_{ij}^{22} & C_{ij}^{23} \\ 
C_{ij}^{31} & C_{ij}^{32} & C_{ij}^{33}    
\end{array}} 
\right) 
= 
\left( 
{\begin{array}{ccc}
0.004 & 0.006 & 0.002 \\       
0.006 & 0.005 & 0.002 \\ 
0.001 & 0.007 & 0    
\end{array}} 
\right)
\end{eqnarray*}
whenever the cells $C_{hl},$ $h,l=1,2,3$ belong to $N_{1}(i,j),$ i.e., for fixed $h$ and $l,$ the coupling strengths $C^{hl}_{ij},$ $i,j=1,2,3,$ are taken to be equal to each other in cases where $C_{hl}$ belongs to $N_{1}(i,j).$

In (\ref{ex1}), for each $i$ and $j,$ we take the rectangular input currents as $P_{ij}(t,\zeta)=\zeta_k,$ $t\in (\theta_k,\theta_{k+1}],$ $k\in \mathbb Z.$ Here, the sequence $\zeta=\left\{\zeta_k\right\}_{k\in\mathbb Z}$ is a solution the logistic map
\begin{eqnarray} \label{logistic}
\zeta_{k+1}=F_{\mu}(\zeta_k),
\end{eqnarray}
where $F_{\mu}(s)=\mu s(1-s).$ For the values of the parameter $\mu\in(0,4],$ the unit interval $[0,1]$ is invariant under the iterations of (\ref{logistic}) \cite{Hale91}. The inverses of the function $F_{\mu}$ with ranges on the intervals $[0,1/2]$ and $[1/2,1]$ are
$
G_{\mu}(s)=\displaystyle \frac{1}{2} \left( 1-\sqrt{1-\frac{4s}{\mu}} \right)
$
and
$
H_{\mu}(s)=\displaystyle \frac{1}{2} \left( 1+\sqrt{1-\frac{4s}{\mu}} \right),
$
respectively.

 One can calculate that 
\[ \sum_{C_{hl} \in N_1(1,1)} C_{11}^{hl} = 0.021,\, \sum_{C_{hl} \in N_1(1,2)} C_{12}^{hl} =  0.025,\, \sum_{C_{hl} \in N_1(1,3)} C_{13}^{hl} = 0.015,\]
\[ \sum_{C_{hl} \in N_1(2,1)} C_{21}^{hl} = 0.029, \, \sum_{C_{hl} \in N_1(2,2)} C_{22}^{hl} = 0.033, \, \sum_{C_{hl} \in N_1(2,3)} C_{23}^{hl} = 0.022,\]
\[ \sum_{C_{hl} \in N_1(3,1)} C_{31}^{hl} = 0.019, \, \sum_{C_{hl} \in N_1(3,2)} C_{32}^{hl} = 0.021,\, \sum_{C_{hl} \in N_1(3,3)} C_{33}^{hl} = 0.014.\]  
Conditions $(C1)-(C3)$ are satisfied for SICNN (\ref{ex1}) with $\gamma=0.3,$ $\delta=0.33,$ $L_f=\sqrt[4]{27}/18,$ $M_f=\pi/18,$ $M_{11}=1,$ $M_{12}=0.6,$ $M_{13}=0.8,$ $M_{21}=0.7,$ $M_{22}=0.3,$ $M_{23}=0.8,$ $M_{31}=0.9,$ $M_{32}=0.8,$ $M_{33}=0.7,$ and $K_0=5.0979.$ Using the results of \cite{Akh8} one can confirm that for each periodic sequence $\zeta=\left\{\zeta_k\right\}_{k\in\mathbb Z},$ SICNN (\ref{ex1}) possesses a unique quasi-periodic output.

\begin{figure}[ht] 
\centering
\includegraphics[width=13.0cm]{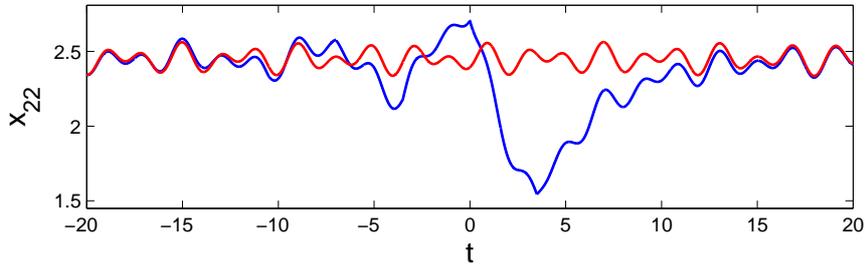}
\caption{\footnotesize Homoclinic output of SICNN (\ref{ex1}). The outputs $\phi^{22}_{\eta}(t)$ and $\phi^{22}_{\zeta^*}(t)$ corresponding to the cell $C_{22}$ are represented in blue and red colors, respectively. The simulation confirms that $\phi_{\eta}(t)$ is homoclinic to the quasi-periodic output $\phi_{\zeta^{*}}(t).$}
\label{fig1}
\end{figure}

In order to demonstrate the existence of a homoclinic motion in the dynamics of (\ref{ex1}), let us take $\mu=3.9.$ The orbit $\eta=\left\{\ldots, H^3_{3.9}(\eta_0), H_{3.9}^2(\eta_0), H_{3.9}(\eta_0), \eta_0, F_{3.9}(\eta_0), F^2_{3.9}(\eta_0), F^3_{3.9}(\eta_0), \ldots \right\},$ where $\eta_0=1/3.9,$ is homoclinic to the fixed point $\zeta^{*}=2.9/3.9$ of (\ref{logistic}) \cite{Avrutin15}.  Let us denote by $\phi_{\eta}(t)=\big\{ \phi^{ij}_{\eta}(t) \big\}$ and $\phi_{\zeta^*}(t)=\big\{ \phi^{ij}_{\zeta^*}(t) \big\}$ the bounded solutions of SICNN (\ref{ex1}) corresponding to $\eta$ and $\zeta^*,$ respectively. According to Theorem \ref{main_theorem}, the output $\phi_{\eta}(t)$ is homoclinic to the quasi-periodic output $\phi_{\zeta^*}(t).$ The graphs of the outputs $\phi^{22}_{\eta}(t)$ and $\phi^{22}_{\zeta^*}(t)$ corresponding to the cell $C_{22}$ of (\ref{ex1}) are represented in Figure \ref{fig1}. The output $\phi^{22}_{\eta}(t)$ is in blue color, whereas $\phi^{22}_{\zeta^*}(t)$ is in red color. The outputs corresponding to the remaining cells, which are not just pictured here, exhibit similar homoclinic behavior. Figure \ref{fig1} manifests that $\left\| \phi_{\eta}(t)-\phi_{\zeta^{*}}(t) \right\| \to 0$ as $t \to \pm \infty,$ i.e., a homoclinic motion takes place in the dynamics of (\ref{ex1}).

\begin{figure}[ht] 
\centering
\includegraphics[width=13.0cm]{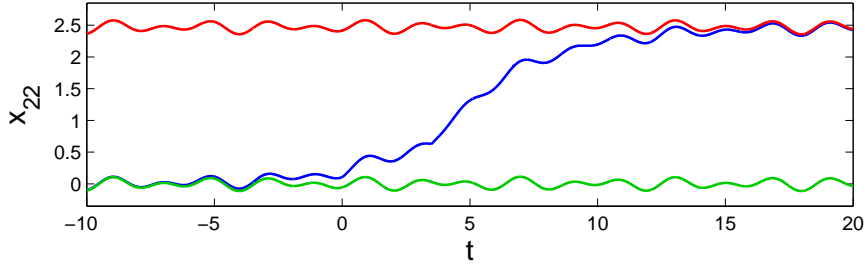}
\caption{\footnotesize Heteroclinic output of SICNN (\ref{ex1}). The outputs $\phi^{22}_{\widetilde{\eta}}(t),$ $\phi^{22}_{\zeta^1}(t)$ and $\phi^{22}_{\zeta^2}(t)$ corresponding to the cell $C_{22}$ are represented in blue, red and green colors, respectively. It is seen in the figure that $\phi_{\widetilde{\eta}}(t)$ is heteroclinic to the quasi-periodic outputs $\phi_{\zeta^{1}}(t)$ and $\phi_{\zeta^{2}}(t).$}
\label{fig2}
\end{figure}

Now, we will demonstrate the presence of a heteroclinic motion of (\ref{ex1}). For that purpose, let us take into account the network (\ref{ex1}) with the value of the parameter $\mu=4.$ It was shown in \cite{Avrutin15} that the orbit $\widetilde{\eta}=\left\{\ldots, G^3_{4}(\widetilde{\eta}_0), G_{4}^2(\widetilde{\eta}_0), G_{4}(\widetilde{\eta}_0), \widetilde{\eta}_0, F_{4}(\widetilde{\eta}_0), F^2_{4}(\widetilde{\eta}_0), F^3_{4}(\widetilde{\eta}_0), \ldots \right\},$ where $\widetilde{\eta}_0=1/4,$ is heteroclinic to the fixed points $\zeta^1=3/4$ and $\zeta^2=0$ of the logistic map (\ref{logistic}). Let $\phi_{\widetilde{\eta}}(t)=\big\{\phi^{ij}_{\widetilde{\eta}}(t)\big\},$ $\phi_{\zeta^1}(t)=\big\{\phi^{ij}_{\zeta^1}(t)\big\}$ and $\phi_{\zeta^2}(t)=\big\{\phi^{ij}_{\zeta^2}(t)\big\}$ be the bounded solutions of SICNN (\ref{ex1}) corresponding to the orbits $\widetilde{\eta},$ $\zeta^1$ and $\zeta^2,$ respectively. It is worth noting that $\phi_{\zeta^1}(t)$ and $\phi_{\zeta^2}(t)$ are quasi-periodic. According to Theorem \ref{main_theorem}, $\phi_{\widetilde{\eta}}(t)$ is heteroclinic to the solutions $\phi_{\zeta^1}(t)$ and $\phi_{\zeta^2}(t).$ The graphs of $\phi^{22}_{\widetilde{\eta}}(t),$ $\phi^{22}_{\zeta^1}(t)$ and $\phi^{22}_{\zeta^2}(t)$ are depicted in Figure \ref{fig2} in blue, red and green colors, respectively. 
Figure \ref{fig2} supports the result of Theorem \ref{main_theorem} such that SICNN (\ref{ex1}) possesses a heteroclinic motion.

\section{Conclusions} \label{SICNNhom_conc}

In this paper, we theoretically prove the existence of homoclinic and heteroclinic outputs in SICNNs whose external inputs are generated by a discrete map. Continuous as well as discontinuous inputs are utilized in the model under investigation. The definitions of homoclinic and heteroclinic motions for the multidimensional dynamics of SICNNs are provided in the functional sense. The presented example reveals that quasi-periodic as well as homoclinic and heteroclinic motions can coexist in the dynamics of SICNNs with discontinuous external inputs. An advantage of the homoclinic and heteroclinic motions obtained in our study is that they can be simulated numerically. Our results can be easily adapted to other types of recurrent neural networks such as Hopfield \cite{Hopfield84} and Cohen-Grossberg \cite{Cohen93} neural networks, and the presented technique may be useful for the mathematical investigation of chaotic behavior in brain dynamics and the rest of the nervous system.


\end{document}